\documentclass{article}

\usepackage{graphicx}

\usepackage[utf8]{inputenc}
\usepackage{amsfonts,amsmath}
\usepackage{todonotes}
\usepackage{color,xcolor}
\usepackage{xspace}
\usepackage{booktabs}
\usepackage{url}
\usepackage{arydshln}
\usepackage{amsmath}
\usepackage{amssymb}
\usepackage{ulem}
\usepackage{lscape}
\usepackage{comment}
\usepackage{float}
\usepackage[]{hyperref}
\usepackage{authblk}

\newcommand{\eg}{{\it e.g.}\xspace}

\newcommand{\p}{\mathsf{p}}

\newcommand{\MFS}{\texttt{MFS}\xspace}
\newcommand{\LFS}{\texttt{LFS}\xspace}

\newcommand{\blue}{\textcolor{black}}

\newtheorem{theorem}{Theorem}

\begin{document}


\title{Practical evaluation of Lyndon factors via\\ alphabet reordering}

\author{Marcelo K. Albertini}
\author{Felipe A. Louza\footnote{louza@ufu.br (Corresponding author).}}

\affil{
Universidade Federal de Uberl\^andia, Uberl\^andia, Brazil
}





\maketitle

\begin{abstract}
\noindent
We evaluate the influence of different alphabet orderings on the Lyndon factorization of a string.
Experiments with \textsc{Pizza\&Chili} datasets show that for most alphabet reorderings, the number of Lyndon factors is usually small, and the length of the longest Lyndon factor can be as large as the input string, which is unfavorable for algorithms and indexes that depend on the number of Lyndon factors. 
We present \blue{results} with randomized alphabet permutations \blue{that can be used as a baseline to} assess the effectiveness of heuristics and methods designed to \blue{modify} the Lyndon factorization of a string via alphabet reordering.
\end{abstract}

\section{Introduction}

A non-empty primitive string $s$ of length $n$ over an ordered alphabet $\Sigma$ is a Lyndon word if $s$ is smaller than all its proper suffixes in alphabetic order, or alternatively, smaller than all its conjugates~\cite{Smyth2003}.
The Lyndon factorization~\cite{ChenFoxLyndon1958} \blue{partitions a string $s$} uniquely into substrings (factors) $w_1, w_2, \dots, w_k$ such that $s=w_1 \cdot w_2\cdots w_k$, and each $w_i$ is a Lyndon word with $w_1 \ge w_2 \ge \dots \ge w_k$.

Lyndon factorization has important applications in combinatorics and string algorithms (see~\cite{Lothaire1997,Smyth2003,Bona2015}).
In the context of suffix sorting, the relative order of suffixes of \blue{a string} inside each Lyndon factor $w_i$ is the same as their order in the original string~\cite{Mantaci2014}, which allows the design of divide-and-conquer algorithms (\eg~\cite{Mantaci2014,Sunita2018}) for computing the suffix array~\cite{Manber1993}. 
Lyndon factors can also be used to compute indexes (\eg~\cite{BannaiKKP19}) based on the Bijective Burrows-Wheeler transform~\cite{Kufleitner09}.
The efficiency of these algorithms, however, depends on the number and length of the Lyndon factors and may deteriorate when the longest factor is large~\cite{Mantaci2014}.
\blue{In Section~\ref{s:exp},} we evaluate the Lyndon factorization of strings from \textsc{Pizza\&Chili} datasets~\cite{PizzaChili} and we show in Table~\ref{t:pizza} that this unfavorable situation is common in practice.

Although the Lyndon factorization of a string $s$ is fixed and unique, it depends on the given alphabet ordering.
Recent approaches (\eg~\cite{Clare2019a,Clare2019b,MajorCDMGZ20}) have investigated how to modify (\blue{increase or reduce}) the number of Lyndon factors via reordering the symbols in the alphabet (considering a new alphabetic order for string comparisons). 
Very recently, Gibney and Thankachan \cite{Gibney2021} showed that the problem of finding an alphabet ordering to either minimize or maximize the number of factors is NP-complete.

In this paper, we 
\blue{evaluate in practice different alphabet orderings to modify the Lyndon factorization of a string}.

In Section \ref{s:alternatives}, we present \blue{two variants of} a \blue{(very)} simple heuristic for reordering the alphabet based on the most frequent symbols of the input string.
In Section~\ref{s:exp},
we compare our heuristic with results by Clare and Daykin~\cite{Clare2019a} and we show that, although simpler, ours produces results close to those presented by Clare and Daykin's method in most cases (even though not satisfactorily, as we discuss next).
Unfortunately, we were not able to execute methods based on evolutionary search techniques presented in~\cite{Clare2019b,MajorCDMGZ20} due their high running time complexity.

In Section~\ref{s:exp}, we conclude that no method consistently modified the Lyndon factors via alphabet reordering \blue{in practice}.
Additionally, we generate uniform random distributions of alphabetic orderings to evaluate their Lyndon factorizations.
Based on these results, 
\blue{we provide a baseline that can be used}
to assess the effectiveness of heuristics and optimization methods that aim to \blue{modify} the number of Lyndon factors and the length of the longest Lyndon factor via alphabet reordering.

\section{Background}\label{s:background}
            
Let $s=s[1]s[2]\dots s[n]$ be a string of size $|s|=n$ over an alphabet $\Sigma=\{\alpha_1, \alpha_2,\dots,\alpha_{\sigma}\}$, with the ordering $\alpha_1 < \alpha_2 < \dots <\alpha_{\sigma}$, \blue{ and $\sigma = |\Sigma|$}.
We denote the set of all strings of symbols in $\Sigma$ by $\Sigma^{*}$.
A substring of $s$ is defined as $s[i,j] = s[i] \dots s[j]$, with $1 \leq i < j \leq n$.
In particular, the substring $s[1, i]$ is a prefix and $s[i, n]$ is a suffix of $s$.
A prefix or suffix of $s$ is called proper if it is not equal to $s$.
We say that $u \in \Sigma^{*}$ is a factor of $s$ if $u$ is equal to some substring $s[i,j]$.

A string $s \in \Sigma^{*}$ is smaller or equal than another string $v \in \Sigma^{*}$, denoted by $s\leq v$, if either $s$ is a prefix of $v$ or $s=w\alpha{}z_1$ and $v=w\beta{}z_2$ with $w, z_1, z_2 \in \Sigma^{*}$ (possibly empty) and the symbols $\alpha,\beta \in \Sigma$, with $\alpha<\beta$.

We denote the repeated concatenation of a string $u$, \blue{$k>1$} times, by $u^k$.
A string $u$ is a repetition if there exists a string $v$ and some integer $k>1$ such that $u=v^k$, otherwise $u$ is primitive.
If a string is primitive, all its conjugates (circular rotations) are distinct.

A non-empty primitive string $s$ is a Lyndon word if it is lexicographically smaller than all its proper suffixes~\cite{ChenFoxLyndon1958}, or smaller than all its conjugates. 
For example, $s=abcacb$ is a Lyndon word.
A Lyndon factor of $s$ is a factor $w$ that is a Lyndon word itself. 

\begin{theorem}[Lyndon Factorization]
Any string $s \in \Sigma^{*}$ has a unique factorization $s=w_1w_2\dots w_k$, such that $w_1 \ge w_2 \ge \dots \ge w_k$ is a non-increasing sequence of Lyndon words.
\end{theorem}

Herein, we denote the number of Lyndon factors as $k$, and the length of the longest factor as $m = \max_{1\leq i \leq k}(|w_i|)$.
The Lyndon factorization can be computed in linear time by using Duval's algorithm \cite{Duval1983}.

For example, given $s = alohomora$ over $\Sigma=\{a,h,l,m,o,r\}$, its Lyndon factorization is \[alohomor \cdot  a\]
with $k=2$ Lyndon factors and longest factor length $m=8$.

It is easy to see that\blue{, for an alphabet $\Sigma = \{a,b\}$, with $\sigma=2$,} the minimum number of Lyndon factors for a string $s$ of length $n$ is $k=1$ with $m=n$, which is \blue{achieved} by \blue{$s=a^{n-1}b$},
while the maximum number of Lyndon factors is $k=n$ with $m=1$, achieved by \blue{$s=ba^{n-1}$}.
In this paper we are interested in evaluate these values for long text datasets \blue{and ASCII alphabets}.

\section{Reordering the alphabet}\label{s:alternatives}

The problem of reordering (permuting) the symbols of a given alphabet $\Sigma=\{\alpha_1, \alpha_2,\dots,\alpha_{\sigma}\}$, with the original ordering $\alpha_1 < \alpha_2 < \dots <\alpha_{\sigma}$, is to create another alphabet $\Sigma_{\pi} = \{ \alpha_{\pi[1]}, \alpha_{\pi[2]}, \dots, \alpha_{\pi[\sigma]}\}$ with the same $\sigma$ symbols of $\Sigma$ permuted in a different order, such as $\alpha_{\pi[1]}  <\alpha_{\pi[2]} < \dots < \alpha_{\pi[\sigma]}$.

As a consequence, two strings in $\Sigma^{*}$ may have different lexicographic orders when considering them from $\Sigma_{\pi}^{*}$.
For example, given the strings $u=aloho$ and $v=mora$, we have that $u<v$ when $u,v\in \Sigma^{*}$, while $u>v$ when $u,v\in \Sigma_{\pi}^{*}$, with $\Sigma_{\pi} = \{r, o, m, l, h, a\}$\blue{, where $r < o < m < l < h < a$}.

The Lyndon factorization of $s$ depends on the ordering of symbols in $\Sigma$.
Therefore, Lyndon factorizations can differ when strings are drawn from different alphabet orderings.

For example, suppose that $s = alohomora$ was written over the alphabet permutation \blue{$\Sigma_{\pi}= \{r, o, m, l, h, a\} $}. 
The Lyndon factorization of $s \in \Sigma_{\pi}^*$ is
\[\blue{a \cdot l \cdot oh \cdot om \cdot o \cdot ra }\]
with \blue{$k=6$} and \blue{$m=2$}.

Whereas another possible alphabet ordering \blue{$\Sigma_{\pi'}= \{m, r, a, h, l, o\}$}, 
would result the following Lyndon factorization for $s \in \Sigma_{\pi'}^*$
\[aloho \cdot mora \]
with $k=2$ and $m=5$

A natural question is whether it is possible to design a efficient method for reordering (permuting) the alphabet to modify the Lyndon factorization, increasing or decreasing $k$ and $m$ based on the input string $s$.
For small alphabets, like DNA, the number of possible choices is small $4!=24$, but for larger alphabets, like ASCII, such brute-force approach becomes unfeasible.

This problem has been recently investigated in~\cite{Clare2019a,Clare2019b,MajorCDMGZ20} with greedy and evolutionary algorithms \blue{(on short sequences in the range of thousands of symbols)}. 
In the next section we present a straightforward heuristic for reordering the alphabet based on frequencies of symbols in $s$, which produce similar results to those by  Clare and Daykin's method~\cite{Clare2019a}.

\subsection{Most/Least Frequent Symbol}\label{s:mfs}

Given the Parikh vector, $\p(s)$, for the string $s$, where $\p(s)$ gives the number of occurrences of each $\alpha_i \in \Sigma$ in $s$.
The most frequent symbol (\MFS) method assigns to $\pi[i]$ the $i$-th most frequent symbol in $\p(s)$, whereas the least frequent symbol (\LFS) method, assigns to $\pi[i]$ the $i$-th least frequent symbol in $\p(s)$. 

\blue{A} new alphabet ordering $\Sigma_{\pi} = \{ \alpha_{\pi[1]}, \alpha_{\pi[2]}, \dots, \alpha_{\pi[\sigma]}\}$ is created accordingly, such as $\alpha_{\pi[1]}  <\alpha_{\pi[2]} < \dots < \alpha_{\pi[\sigma]}$.

In the case different symbols have the same values in $\p(s)$, we consider their original ranks in $\Sigma$.

For example, given $s = alohomora$ over $\Sigma=\{a,h,l,m,o,r\}$.
We have
$\p(s) = [2, 1, 1, 1, 3, 1]$
that give us the following alphabets reorderings:
\[\Sigma_{\MFS} = \{o,a,h,l,m,r\} \mbox{ and } \Sigma_{\LFS} = \{h,l,m,r,a,o\}\]

The Lyndon factorization of $s \in \Sigma_{\MFS}^*$ is
\[al \cdot ohomora \]
with $k=2$ and $m=7$, whereas the factorization of $s \in \Sigma_{\LFS}^*$ is 
\[a \cdot lo \cdot homora \]
with $k=3$ and $m=6$.


\section{Experiments}\label{s:exp}


\begin{table}[!t]
\renewcommand{\arraystretch}{1.5}
\centering
\caption{
Experiments with \textsc{Pizza\&Chili} datasets. 
The datasets {\tt einstein-de}, {\tt kernel}, {\tt fib41} and {\tt cere} are highly repetitive texts. 
The dataset {\tt english.1G} is the first 1GB of the original english dataset.
}
\begin{tabular}{l|l|r|c|c}
dataset     & $\sigma$ & {size in MB} & {number of factors} & {longest factor} \\ \hline
\texttt{sources    } & 230      & 201    & 30   & 52.00\%    \\
\texttt{dblp       } & 97       & 282    & 17   & 37.93\%    \\
\texttt{dna        } & 16       & 385    & 18   & 74.75\%    \\
\texttt{english.1GB} & 239      & 1,047  & 30   & 57.28\%    \\
\texttt{proteins   } & 27       & 1,129  & 24   & 80.71\%    \\ \hdashline
\texttt{einstein-de} & 117      & 88     & 44   & 40.39\%    \\
\texttt{kernel     } & 160      & 246    & 33   & 41.73\%    \\
\texttt{fib41      } & 2        & 256    & 21   & 61.68\%    \\ 
\texttt{cere       } & 5        & 440    & 22   & 79.98\%  \\ 
\bottomrule
\end{tabular}
\label{t:pizza}
\end{table}


We evaluated the Lyndon factorization of strings from \textsc{Pizza\&Chili}~\cite{PizzaChili} datasets.
The source-code of methods \MFS and \LFS (Section~\ref{s:mfs}) is freely available at \url{https://github.com/felipelouza/remap/}.
Table~\ref{t:pizza} shows 
the alphabet size (Column 2),
string length in MB (Column 3),
number of Lyndon factors (Column 4) and 
the longest Lyndon factor length  in percentage to the string length (Column 5) of each dataset.
We considered the standard alphabet ordering in this first experiment.

Notice that the number of Lyndon factors is small even for larger strings ({\tt english.1GB} and {\tt proteins}) and the longest Lyndon factor can be as large as the input string.
This situation is particularly unfavorable for divide-and-conquer algorithms that take advantage of the input string partitioned into Lyndon factors to compute fundamental data structures for string processing (\eg~\cite{Mantaci2014,Sunita2018}).

\subsection{Alphabet reordering}

We compared the methods \MFS and \LFS (Section~\ref{s:mfs}) with the {\tt greedy} algorithm (with and without backtracking) proposed in~\cite{Clare2019a}.
Unfortunately, we were not able to run the evolutionary methods proposed in~\cite{Clare2019b,MajorCDMGZ20} due their high time complexity (the authors presented experiments only with very small inputs).
We also included results obtained from 100 samples drawn from a uniform distribution of permutations. 
These {\tt random} results are presented as box plots in the figures.

\begin{figure}[t]
    \centering
    \includegraphics[width=1\textwidth]{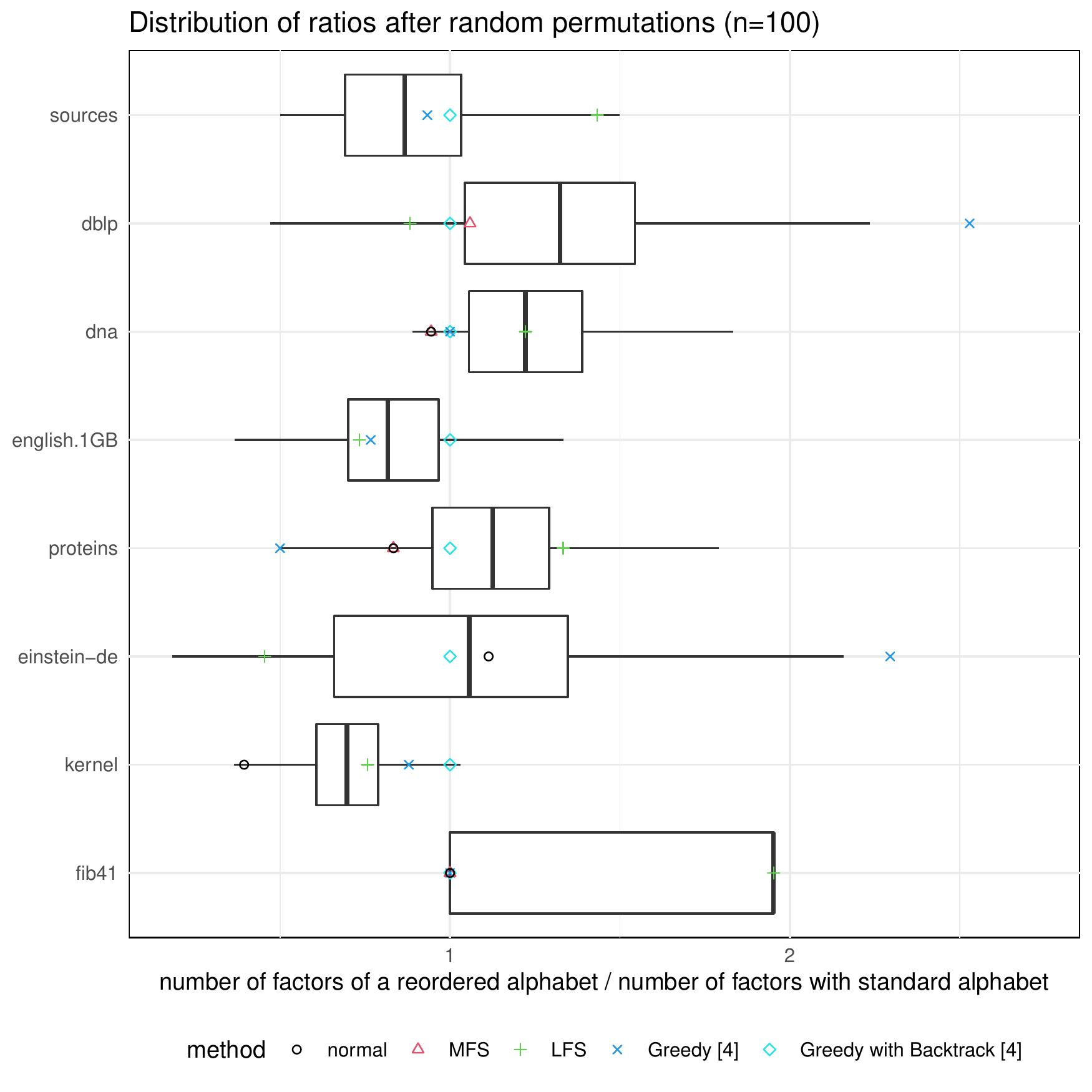} 
    \caption{Number of factors for each method. 
    Results for dataset {\tt cere} were omitted due to its much larger scale.
    }
    \label{fig:numfactors}
\end{figure}

Figure \ref{fig:numfactors} shows the number of Lyndon factors resulting for each method compared to the number of factors presented in Table~\ref{t:pizza} (standard alphabet ordering).
Results for dataset {\tt cere} were omitted, \MFS and {\tt random} generated more than 7,000 factors, while \LFS, {\tt greedy} and {\tt greedy with backtracking} generated about 2,000 factors.
Despite this case, the resulting number of factors is still small for all alphabet orderings, 
\blue{at most twice the number created by the original alphabet order.}

\begin{figure}[t]
    \centering
    \includegraphics[width=1\textwidth]{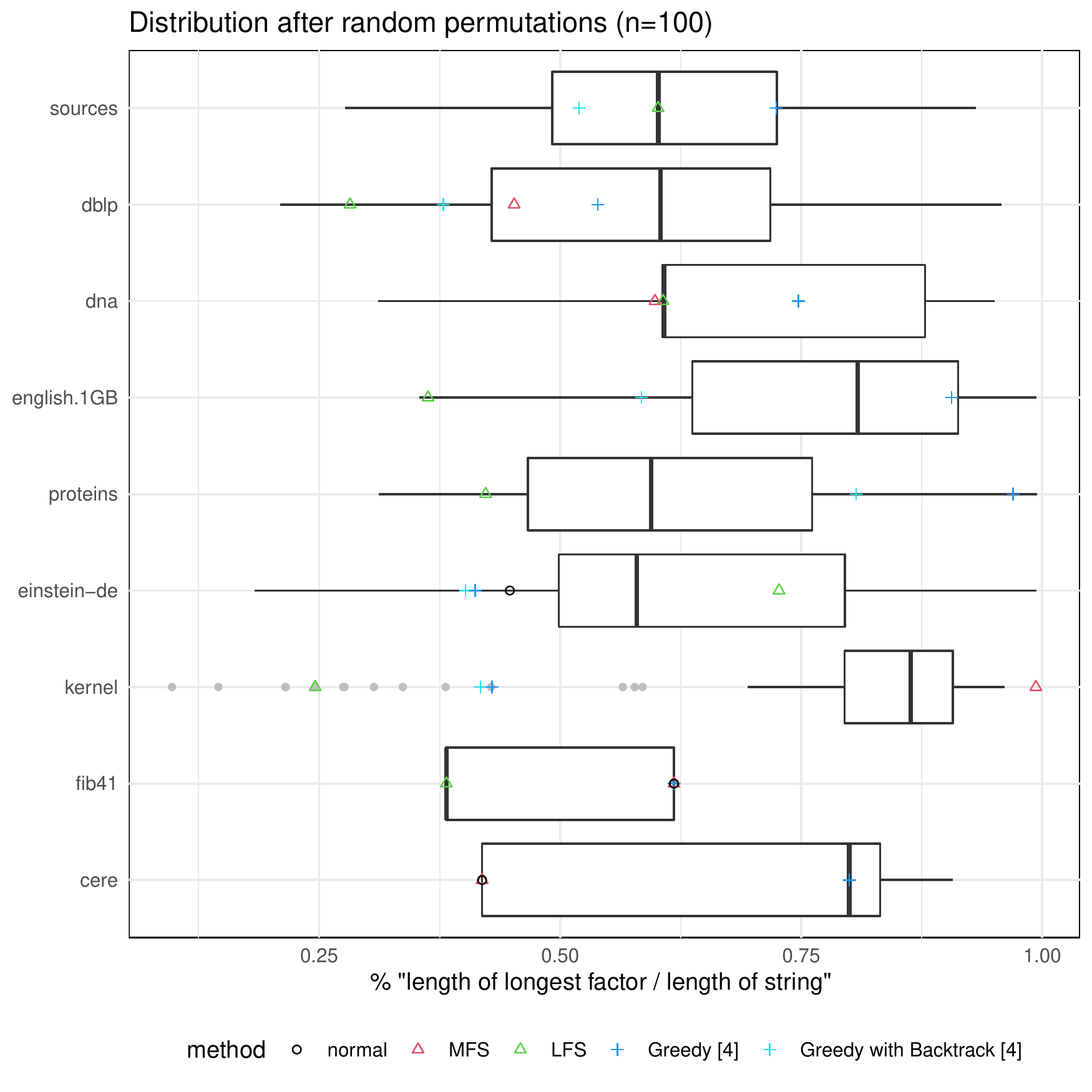}
    \caption{Longest Lyndon factor length for each method. Grey points are outliers in the random distributions.
    }
    \label{fig:maxfactors}
\end{figure}

Figure \ref{fig:maxfactors} shows the lengths of the longest Lyndon factors in percentage to the total length of the input string for each method.
Notice that, the length of the longest factor can be as large as the length of the input string.
Note that in most cases, the longest factor is not smaller than $33\%$ of the input string.
Also, results of the straightforward methods \MFS and \LFS were close to the {\tt greedy} algorithms proposed by Clare and Daykin~\cite{Clare2019a}.

We remark that no method consistently always increased (or decreased) neither the number of Lyndon factors nor their maximum lengths.

\subsection{\blue{Randomized alphabet}}

Tables \ref{t:random_factors} and \ref{t:random_longest} \blue{summarizes} results of the 100 samples of randomized alphabet permutations (method {\tt random}).
We \blue{present a} \blue{baseline that can be used}
to assess heuristics and optimization methods for increasing/decreasing Lyndon factors via alphabet reordering based on these results.
\blue{The histograms of these results can be found in the appendix.}

We state that \blue{any} maximization method can be considered effective if it consistently selects an alphabet permutation $\Sigma_{\pi}$ that results into a number of Lyndon factors (or the longest factor length) larger than most other random permutations in Q3 (third quartile, $> 75\% $).
Similarly, a minimization method will be effective if the selected alphabet permutation provides a number of Lyndon factors (or the longest factor length) smaller than most other permutations in Q1 (first quartile, $< 25\%$).

\begin{table}[t]
\renewcommand{\arraystretch}{1.5}
\centering
\caption{
Number of Lyndon factors for random permutations of the alphabet.
Column \blue{2} shows the number of factors for the original alphabet.
Columns \blue{3} and 7 show the minimum and maximum number of factors obtained with 100 random permutations of the alphabet.
Columns \blue{4, 5 and 6} show the first quartile (Q1), the median and the third quartile (Q3) of the distribution.
}
\begin{tabular}{l|c|ccccc}
\texttt{}            &       \textbf{original}        &     \textbf{min}        & \textbf{Q1}            &
\textbf{median}         &\textbf{Q3}             & \textbf{max}            \\ \hline
\texttt{sources}     & 30          & 15          & 21          & 26          & 31          & 54          \\
\texttt{dblp}        & 17          & 13          & 17          & 21          & 25          & 43          \\
\texttt{dna}         & 18          & 17          & 20          & 23          & 24          & 35          \\
\texttt{english.1GB} &  20          & 11           & 7           & 13          & 19          & 28          \\
\texttt{proteins}    & 24          & 13          & 22          & 25          & 29          & 43          \\ \hdashline
\texttt{einstein-de} & 44          & 21          & 43          & 57          & 71          & 123         \\ 
\texttt{kernel}      & 55          & 10          & 18          & 22          & 28          & 39          \\
\texttt{fib41}       & 21          & 22          & 22          & 22          & 41          & 41          \\
\texttt{cere}        & 22          & 16          & 2,192       & 7,718       & 7,724       & 7,728       \\ 
\bottomrule
\end{tabular}
\label{t:random_factors}
\end{table}

\begin{table}[!t]
\renewcommand{\arraystretch}{1.5}
\centering
\caption{
Longest Lyndon factors (in percentage to the total length of the string) for random permutations of the alphabet.
Column 1 shows the longest factor of the original alphabet.
Columns 3 and 7 show the minimum and maximum longest factors obtained with 100 random permutations of the alphabet.
Columns 4, 5 and 6 show the first quartile (Q1), the median and the third quartile (Q3) of the distribution.
}
\begin{tabular}{l|c|ccccc}
\texttt{}            &       \blue{\textbf{original}}        &     \textbf{min}        & \textbf{Q1}            &
\textbf{median}         &\textbf{Q3}             & \textbf{max}            \\ \hline
\texttt{sources}     & {52.00\%} & 27.72\% & 49.19\% & 60.20\% & 72.50\% & 93.15\% \\ 
\texttt{dblp}        & {37.93\%} & 20.98\% & 42.96\% & 60.50\% & 71.91\% & 95.91\% \\ 
\texttt{dna}         & {74.75\%} & 31.11\% & 60.68\% & 60.82\% & 87.91\% & 95.11\% \\ 
\texttt{english.1GB} & {57.28\%} & 34.66\% & 62.43\% & 79.23\% & 89.45\% & 97.43\% \\ 
\texttt{proteins}    & {80.71\%} & 31.22\% & 46.65\% & 59.46\% & 76.16\% & 99.46\% \\  \hdashline
\texttt{einstein-de} & {40.39\%} & 18.37\% & 50.11\% & 58.25\% & 79.93\% & 99.93\% \\  
\texttt{kernel}      & {41.73\%} & 9.74\%  & 79.51\% & 86.36\% & 90.74\% & 96.09\% \\ 
\texttt{fib41}       & {61.68\%} & 38.12\% & 38.12\% & 38.12\% & 61.68\% & 61.68\% \\ 
\texttt{cere}        & {79.98\%} & 41.89\% & 41.89\% & 79.98\% & 83.18\% & 90.74\%  \\
\bottomrule
\end{tabular}
\label{t:random_longest}
\end{table}

%

\section*{Code availability}

The source-code is freely available at \url{github.com/felipelouza/remap/}.


%
%
%
%
%

\begin{thebibliography}{10}
\providecommand{\url}[1]{{#1}}
\providecommand{\urlprefix}{URL }
\expandafter\ifx\csname urlstyle\endcsname\relax
  \providecommand{\doi}[1]{DOI~\discretionary{}{}{}#1}\else
  \providecommand{\doi}{DOI~\discretionary{}{}{}\begingroup
  \urlstyle{rm}\Url}\fi

\bibitem{BannaiKKP19}
Bannai, H., K{\"{a}}rkk{\"{a}}inen, J., K{\"{o}}ppl, D., Piatkowski, M.:
  Indexing the bijective {BWT}.
\newblock In: Proc. CPM,
  \emph{LIPIcs}, vol. 128, pp. 17:1--17:14. Schloss Dagstuhl - Leibniz-Zentrum
  f{\"{u}}r Informatik (2019)

\bibitem{Bona2015}
Bona, M.: {Handbook of enumerative combinatorics}.
\newblock Discrete Mathematics and Its Applications. CRC Press, Hoboken, NJ
  (2015)

\bibitem{ChenFoxLyndon1958}
Chen, K.T., Fox, R.H., Lyndon, R.C.: Free differential calculus. {IV} -- the
  quotient groups of the lower central series.
\newblock Ann. Math. \textbf{68}, 81--95 (1958)

\bibitem{Clare2019a}
Clare, A., Daykin, J.W.: {Enhanced string factoring from alphabet orderings}.
\newblock Information Processing Letters \textbf{143}, 4--7 (2019)

\bibitem{Clare2019b}
Clare, A., Mills, T., Daykin, J.W., Zarges, C.: {Evolutionary search techniques
  for the Lyndon factorization of biosequences}.
\newblock pp. 1543--1550 (2019)

\bibitem{Duval1983}
Duval, J.P.: Factorizing words over an ordered alphabet.
\newblock Journal of Algorithms \textbf{4}(4), 363 -- 381 (1983)

\bibitem{PizzaChili}
Ferragina, P., Navarro, G.: {Pizza\&Chili}.
\newblock \url{pizzachili.dcc.uchile.cl/}

\bibitem{Gibney2021}
\blue{
Gibney, D., Thankachan, S.V.: Finding an optimal alphabet ordering for lyndon factorization is hard.
\newblock In: Proc. STACS, \emph{LIPIcs}, vol. 187, pp. 35:1--35:15. Schloss Dagstuhl - Leibniz-Zentrum f{\"{u}}r Informatik (2021)
}

\bibitem{Kufleitner09}
Kufleitner, M.: On bijective variants of the burrows-wheeler transform.
\newblock In: J.~Holub, J.~Zd{\'{a}}rek (eds.) Proc. PSC, pp. 65--79 (2009)

\bibitem{Lothaire1997}
Lothaire, M.: Combinatorics on words, Second Edition.
\newblock Cambridge mathematical library. Cambridge University Press (1997)

\bibitem{MajorCDMGZ20}
Major, L., Clare, A., Daykin, J.W., Mora, B., Gamboa, L.J.P., Zarges, C.:
  Evaluation of a permutation-based evolutionary framework for lyndon
  factorizations.
\newblock In: Proc. {PPSN}, vol. 12269, pp. 390--403. Springer (2020)

\bibitem{Manber1993}
Manber, U., Myers, G.: Suffix arrays: a new method for on-line string searches.
\newblock SIAM Journal on Computing \textbf{22}(5), 935--948 (1993)

\bibitem{Mantaci2014}
Mantaci, S., Restivo, A., Rosone, G., Sciortino, M.: {Suffix array and Lyndon
  factorization of a text}.
\newblock Journal of Discrete Algorithms \textbf{28}, 2--8 (2014)

\bibitem{Smyth2003}
Smyth, W.: Computing patterns in strings.
\newblock Pearson Education (2003)

\bibitem{Sunita2018}
Sunita, Garg, D.: {Extended suffix array construction using Lyndon factors}.
\newblock Sadhana - Academy Proceedings in Engineering Sciences \textbf{43}(8),
  1--9 (2018)

\end{thebibliography}



\clearpage\section*{Appendix}

\begin{figure}[h]
    \centering
    \includegraphics[width=1\textwidth]{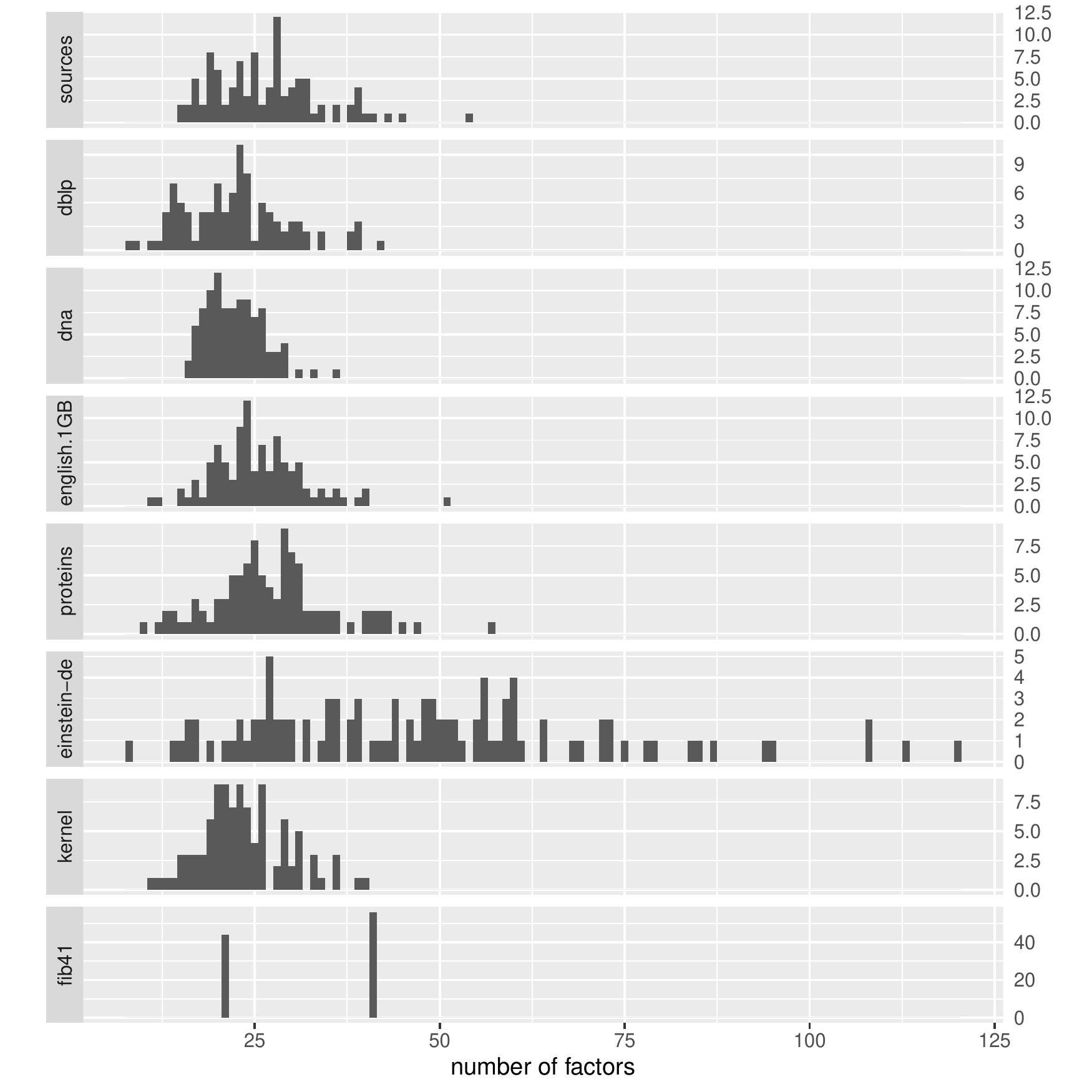} 
    \caption{Number of factors for random permutations of the alphabet. Shapiro-Wilk normality tests, with $p=0.05$, discarded normality of all distributions, except for kernel dataset.
    }
    \label{fig:histograms}
\end{figure}

\begin{figure}[h]
    \centering
    \includegraphics[width=1\textwidth]{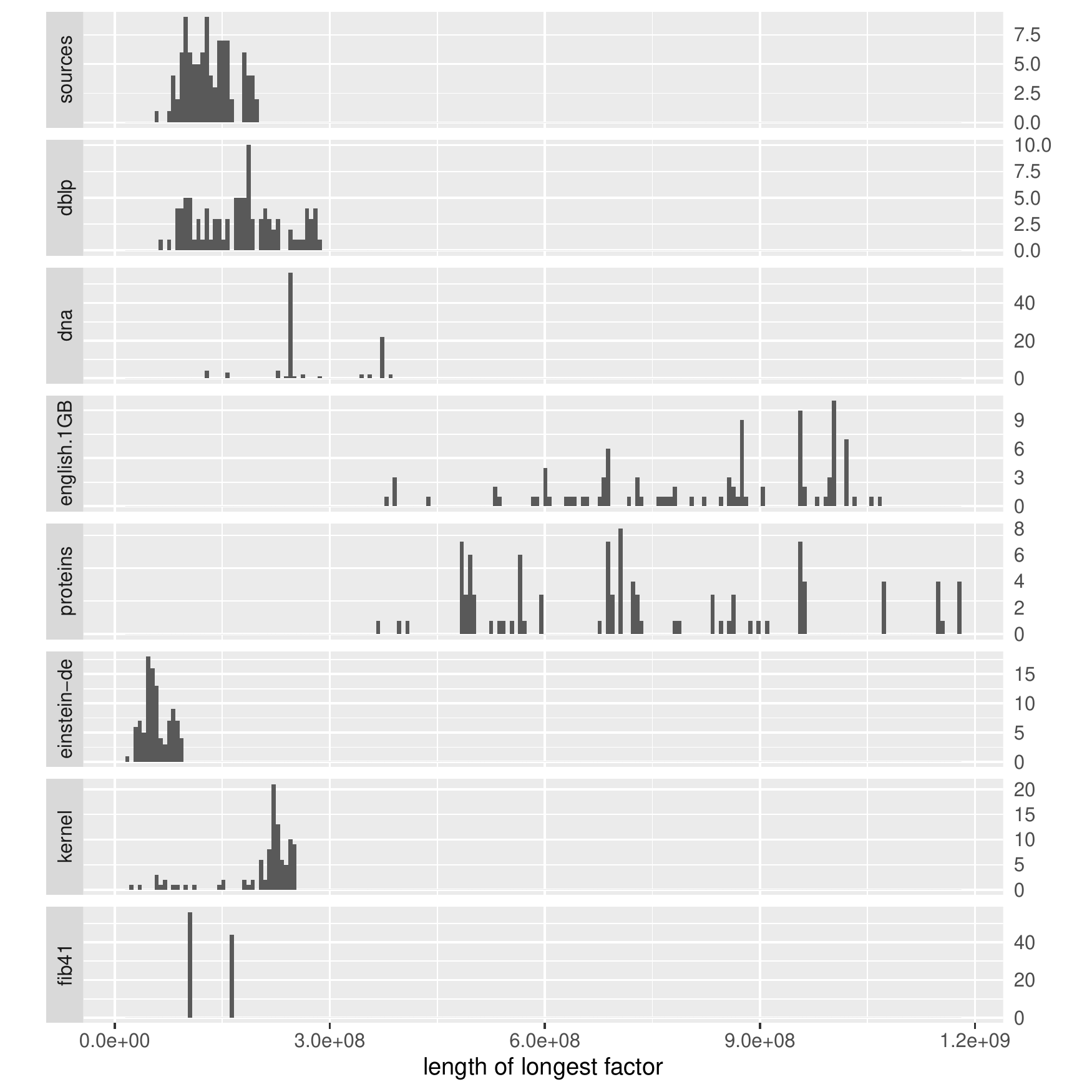} 
    \caption{Longest length of factors for random permutations of the alphabet. Shapiro-Wilk normality tests, with $p=0.05$, discarded normality of all distributions.
    }
    \label{fig:histograms}
\end{figure}

\end{document}